\documentclass[preprint,12pt]{elsarticle} 
\usepackage{graphicx} 
\usepackage{amssymb} 
\usepackage{amsmath} 

\journal{Nuclear Physics A} 

\begin{document} 

\begin{frontmatter} 

\title{Neutron Star News and Puzzles} 
\author{Madappa Prakash} 
\address{Department of Physics and Astronomy, Ohio University, Athens, OH 45701,
USA} 

\begin{abstract} 
Gerry Brown has had the most influence on my career in Physics, and my life after graduate studies.  This article  gives a brief account of some of the many ways in which Gerry shaped my research. 
Focus is placed on the significant strides on neutron star research made by the group at Stony Brook, which Gerry built from scratch. Selected puzzles about neutron stars that remain to be solved are noted. 
\end{abstract}


\end{frontmatter} 

\section{Memories of Gerry Brown (Personal Perspective)}
\label{sec:memroes}  
Wolfram Weise, a long-time collaborator of Gerry's, came up to me after my talk at the meeting and said ``Write exactly as you spoke''. I will try to recollect what I said about Gerry during the meeting.

\subsection{How it all began}
My association with Gerry began in 1980 at the Niels Bohr Institute (NBI) in Copenhagen.  Through Jakob Bondorf, Gerry had 
learned that I was spending most of my meagre 
Danish International Development Agency's  (DANIDA) 
scholarship money calling my wife Manju, who was a physics graduate student in Columbia University, NY.  From the front bench - during a traditional Monday morning seminar at the NBI - Gerry passed to me, in a rear bench,  a yellow sheet of paper.  The first sentence in his note offered me a post-doc position, and the second sentence instructed me to contact the administrator  of the Nuclear Theory Group at  Stony Brook University, Sydel Blumberg. It was  signed, Gerry Brown. I was astounded! 

I later learned  that he was the supervisory editor of Nuclear Physics A and read everything  that was submitted to the journal. I had written a paper with Shalom Shlomo on Wigner distribution functions of nuclei,  and Gerry had chosen Nandor Balaz, an expert on the subject, as the referee of our manuscript. Gerry's comment was ``If you can pass through Nandor, you must be ok.'' Little did I know then that he had plans for me.      
\subsection{My early days with Gerry}
I came to Stony Brook in the summer of 1981 and was put up at the Sunwood estate. I had no car, so I walked everyday to the Physics Department and then back to the ``Golden Cage'' at Sunwood.  Gerry was at the NBI telling Hans Bethe what to work on (supernovae, I later learned). I gave my first talk on ``Phase Space Distributions of Time-Dependent Hartree-Fock (TDHF) Collisons'' to experimentalists in the basement of the Physics building.  My talk, which Aage Bohr and Ben Mottelson had liked at the NBI,  was a disaster at Stony Brook! I thought no one cared. The ever smiling and always kind Linwood Lee was my host, and made me feel that I was doing something useful and that I would be alright at Stony Brook.  

With few people to talk to (post-docs Jochen Wambach, Berndt Schewizinger, and Rudolf Fiebig being busy with their own work), I started a collaboration with John Alexander (of the Chemistry Department) and Rich Friefelder (a physics graduate student doing experiments) on fission fragment angular distributions (of which I knew something from my Ph.D. work). Jochen kindly rented me a room in his house at 295 Sheep Pasture Road, which was less far to the Department than Sunwood. As I was getting somewhat settled, Sydel passed on a message from Gerry that I was to give a talk at MIT and I was to contact John Negele about my visit to MIT. The man was devious! Nervous as I was, I went to MIT, gave my talk, and was not seriously hurt - John was  a kind host.

Gerry returned to Stony Brook in the fall, and was furiously at work on core-collapse supernovae, the little-bag model, Skyrmions, Fermi liquid theory, etc. To me, he said ``you do too much numerical work (TDHF), start thinking!'' He knew how to get under my skin. So I wrote a paper on ``Effective masses in nuclei and the level density parameter'' with Jochen Wambach and Mrs. Ma (a Chinese visitor) using only a calculator. The emphasis was on delineating the roles of the $k$-mass and the $e$-mass. 
I recall Claude Mahaux, editor of Physics Letters B, liking this paper. Gerry just nodded.   
Graduate students Jerry Cooperstein (Coop) and Eddie Baron, professors Jim Lattimer and Amos Yahil, and post-doc Adam Burrows, and Hans Bethe, through his daily faxes, all working on supernova explosions, consumed Gerry's time to a great extent. 
He had me working on sub-threshold pion production, a topical subject at the time, but we were not satisfied with the results (not a  large enough cross section for his taste).   A little later, I  worked with Peter Braun-Munzinger and Johanna Stachel using  a different approach.  We wrote a few papers together and got to know each other very well. Watching the excitement of the supernova gang,  I asked him to give me a supernova-related project to work on.  He said ``You're doing very well on your own;  I don't have to worry about you. Leave me alone.''  End of conversation. So I was just hanging around, helping Coop, Eddie, and Karen Kohlemainen (Lattimer's student) when I could. 

Things turned around for me one day, when  Gerry asked me to come to his house for morning coffee. The entire supernova gang was there. Gerry was fussing around Hans Bethe trying to make him comfortable. Reams of computer outputs from Stan Woosely were on the kitchen-cum-dining room table. Bethe started writing on a stand-up black board and said something about which Adam Burrows from the end of the room said ``That's manifest nonsense!''.  Utter silence in the room. I distinctly recall Gerry's death-stare at Adam.   Bethe paused, said he would be right back, and went to the bath room. We could all hear the toilet flush. He came back and said to Adam ``You're right''. I was ecstatic! I thought, this is the place for me; I could say anything I wanted.  Since then, Adam and I have talked to each other a lot, but never wrote a paper together, although we were toying with the idea of writing a book on ``Fun with Fermi Integrals'', as we both had several analytical results not found easily in the literature at the time. We talked at this meeting about it and decided that time was long past for the book.  

\subsection{Mentor, colleague, in-situ father and a great friend}
For several years, Gerry invited me and my wife to Thanksgiving and Christmas dinners. It was only when I said that I wanted to learn how to cook these dinners myself that we were reluctantly let off. Among others, Peter, Johanna, and I took turns playing tennis doubles with Gerry and his wife Betty. Playing them was no easy task as both of them towered over the net, and Gerry was particularly competitive. He would call me up on New Year's eve and wanted to play tennis at 9 P.M. or so. When I complained that I was having a party at my house, he would say ``Never mind, come over. You'll enjoy your drinks better later.'' So, I went and played. He had given me a new tennis racquet so I could  play him better.  

Gerry  often killed me with his kindness. He would come over with a problem or a new paper, and say ``You seem to know this. Tell me all about it. No rush, tomorrow coffee-time will do'', and just walk away.  I  knew nothing of what he was talking about. The man was devious!  I would study all night, come to work and hide in a corner of the library hoping he would forget. No chance! He always found me no matter where I sat in the library, put his hands on my shoulders from the back and say, ``let's go talk''.  

\subsection{A perfect ``match-maker''} 

I haven't met anyone who could put people together to attack a problem as well as Gerry. He had me working with Kevin Bedell, Jim Lattimer (who became a long-term collaborator), Jean-Paul Blaizot, Wolfram Weise, Manque Rho, Subal Das Gupta, etc., in addition to many students and post-docs (I got to write papers with most of them). The number of people in this audience is a testimony to his ``match-making'' ability, as Eddie has already pointed out. 
Kevin, who lived in the same apartment complex as I did,  taught me tennis and Fermi liquid theory (we wrote a paper together on the ``Incompressibility of neutron-rich matter''), and even sold me his car.  Gerry was also open to suggestions. 
I mentioned to him that it would be nice to have Subal Das Gupta visit us so we could work on the momentum-dependence of the nuclear equation of state to be used in heavy-ion collisions. Subal was invited to spend his sabbatical year at Stony Brook, which proved very productive. Later, Charles Gale, Subal's student, also spent a part of his sabbatical at Stony Brook. 
When I told him that Achim Schwenk would benefit from working with Bengt Friman at GSI, it was a done deal.  

\subsection{Provocateur par excellence} 

Provoking people was Gerry's favorite past time. He would often startle me in my office (you could never tell when he would be coming as he walked silently for such a big man) with ``It's unbelievably beautiful''. When I asked him what was beautiful, he would expound at length on something he had thought up the night before.  I often thought it was outrageous, but always remembered Manque's advise to me, ``Gerry's starting point is always right, his conclusions are always right, for everything in between you're on your own''. It took me quite a while to realize that Gerry was deliberately goading me.  I slowly learned how to deal with it. I had him write it down for me, which he always did. Then, I would work out the problem his way and my way, and give him notes on both approaches. It took him time to abandon his method when wrong, but my system generally worked. 

Knowing fully well that I had played professional cricket in India, he deliberately did not tell me that John Millener at the Brookhaven National Laboratory (BNL) was an excellent cricketer (an Oxford blue), and that BNL had a decent cricket team.  Devious again!  I ended up playing for a gay soft-ball team at Stony Brook. We even reached the finals as no other team would play us because it was a gay team. The news got published in the local newspapers. Gerry's reaction was ``I didn't bring you here from Copenhagen to learn Americana! Get to serious work.'' Eventually, I met John at a Nancy and Ernie Warburton's party, and we played many games together at BNL, Oyster Park, and Staten Island where the ``Carribean Cricketers'' would try to intimidate us.  We won some and lost some, but were not intimidated.  

I once asked Gerry for some guidance on the axial coupling constant $g_A$ in medium, as he had worked on it before. He said 
``My dear boy, every theoretical physicist  worth his salt must write a paper on $g_A$. I've paid my dues to $g_A$. You're on your own''.  He altogether refused to help me. When I handed him my paper with Greg Carter, he said ``Nice''. That was nice!

I know that I lost my temper with Gerry sometimes. But, Gerry was always kind to me. For reasons entirely unknown to me, he believed in me. 
  
\subsection{Generous beyond belief in personal and professional matters}

Genuine concern for all students and post-docs  was always in evidence with Gerry. He helped them financially when needed, often from his own pocket when things could not be arranged otherwise. He talked with them at length about their life and arranged help with his many contacts. As supervisory editor of  Physics Letters B and Nuclear Physics A,  he accepted  many papers for publication even when he, or the referees, disapproved of their contents. His take was ``Let people in the field decide''.   
He gave full credit to my work with Lattimer on what might have happened with SN 1987A. The so-called Brown-Bethe scenario that the neutron star there might have become a black hole was generously credited as the Prakash-Lattimer scenario in his talks on the subject. Being a passionate lover of physics who never stopped doing physics, he encouraged others to be the same. The number of times he called me at home to talk physics is uncountable. He regularly called me late at night to talk physics even when I moved to Ohio. He dearly wanted to write a paper with me on $^{14}{\rm C}$, even after his papers with Jeremy Holt, but unfortunately that never came to pass.

\section{On-going activities spawned by Gerry}
\label{sec:activities} 
Not surprisingly, the research of many of us trained at the Nuclear Theory Group in Stony Brook has centered around topics that were dear to Gerry. Here, I'll highlight the research of Gerry's wards that have made news along with topics that puzzle me and others in the field of neutron stars. 

\subsection{Maximum and minimum masses of neutron stars}
\label{sec:masses}
The discovery of well-measured neutron star masses $1.97\pm0.04~{\rm M}_\odot$ \cite{Demorest10} and 
 $2.01\pm0.04~ {\rm M}_\odot$ \cite{Antoniadis13}  has caused quite a stir among theorists. The cause for excitement is that many, but not all,  of the scenarios in which a softening of the neutron-star matter equation of state (EOS) occurred due to the presence of Bose condensates, hyperons, quark matter, etc., fail to support 2 solar masses.  In their article ``What a two solar mass neutron star really means'', Lattimer and Prakash \cite{LP:11} have examined the implications of high mass neutron stars including an estimate of the $2.4~{\rm M}_\odot$ for the black widow pulsar. Based on the pressure $p$ vs. energy density $\epsilon$ relation (EOS) 
\begin{equation}
 p = 0 \quad {\rm for} \quad \epsilon < \epsilon_0\,; \quad p = \epsilon - \epsilon_0 \quad {\rm for} \quad \epsilon > \epsilon_0\,
 \label{compeos}
\end{equation}
that produces the most compact configuration (that is, the largest value of $GM/c^2R$) \cite{Haensel89,Koranda97} so that limits to the maximum mass and minimum radius could be set, model-independent upper limits to thermodynamic properties in neutron stars which only depend upon the neutron star maximum mass were established from causality considerations. 
The EOS in equation (\ref{compeos}) is at the causal limit as $dp/d\epsilon=(c_s/c)^2=1$, where $c_s$ is the adiabatic speed of sound and contains a single parameter $\epsilon_0$ which features in all physical observables due to scaling relations displayed by the structure (TOV) equations \cite{Witten84}. 

\begin{figure}[hbt]
\vskip -1cm
\begin{center}
{\includegraphics[width=300pt,angle=90]{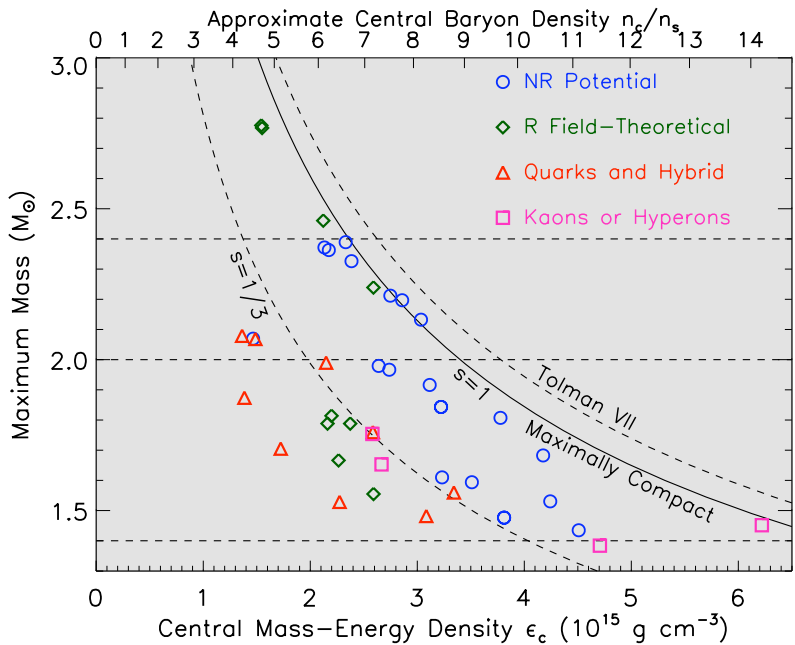}}
\end{center}
\vskip -1.5cm
\caption{(Color online) Maximum mass versus central mass-energy density (bottom $x$-axis) and central baryon density (top $x$-axis) 
for the maximally compact EOS in equation (\ref{compeos}). 
The curve labeled $s=1/3$ corresponds to  $p=(\epsilon-\epsilon_0)/3$ characteristic of commonly used quark matter EOSs.  Results of Tolman VII solution \cite{Tolman39}  with  
$\epsilon=\epsilon_c(1-(r/R)^2)$ and for various model calculations of neutron star matter - see inset for legends - are as shown. Figure adapted from Ref. \cite{LP:05}.}
\label{ultimate}
\end{figure}

Using $2~{\rm M}_\odot$ as a proxy for the true maximum mass, Ref. \cite{LP:11} showed that the energy density cannot exceed about 2 GeV fm$^{-3}$, the pressure about 1.3  GeV fm$^{-3}$,  and the baryon chemical potential about 2.1 GeV.  In the case of self-bound quark matter stars, these limits were reduced to 1.3 GeV fm$^{-3}$,  0.9 GeV fm$^{-3}$,  and 1.5 GeV, respectively.  
Figure \ref{ultimate} shows how severely the limit on the maximum central energy density would be reduced in the case of a well-measured  $2.4~{\rm M}_\odot$ mass. Also shown in this figure are limits from the Tolman-VII solution \cite{Tolman39} that  model EOS's are unable to reach.

%
%
%
%

The limit on the value of the chemical potential deduced for quark matter stars is worthy of comment.  For the most part, calculations of the quark matter EOS have relied on perturbative treatments, the latest being that of Ref. \cite{Kurkela10} in which a complete calculation up to 2-loops was reported.   The values of the chemical potential needed for a perturbative treatment to be valid far exceeds the bound set in Ref. \cite {LP:11}  for self-bound quark stars. In view of this, hybrid stars, those in which quark matter resides only in the core, but is  surrounded by a nuclear matter mantle was undertaken by Alford, Han and Prakash \cite{Alford13}. In this work, generic conditions for stable hybrid stars were established using the EOS
\begin{equation}
\epsilon(p) = \left\{\!
\begin{array}{ll}
\epsilon_{\rm NM}(p) & \quad p<p_{trans}\\
\epsilon_{\rm NM}(p_{trans})+\Delta\epsilon+c_{\rm QM}^{-2} (p-p_{trans}) & \quad p>p_{trans}
\end{array}
\right.
\label{eqn:EoSqm1}
\end{equation}
where $\epsilon_{\rm NM}(p)$ denotes the nuclear matter EOS, $\Delta\epsilon$ is the discontinuity in energy density $\epsilon$ at the transition pressure $p_{trans}$, and $c_{QM}^2$ is the squared speed of sound of quark matter taken constant with density but varied in the range 1/3 (characteristic of perturbative quark matter) to 1 (causal limit).    

%
\begin{figure}[thb]
\vskip -0.25cm
{\includegraphics[width=200pt,angle=0]{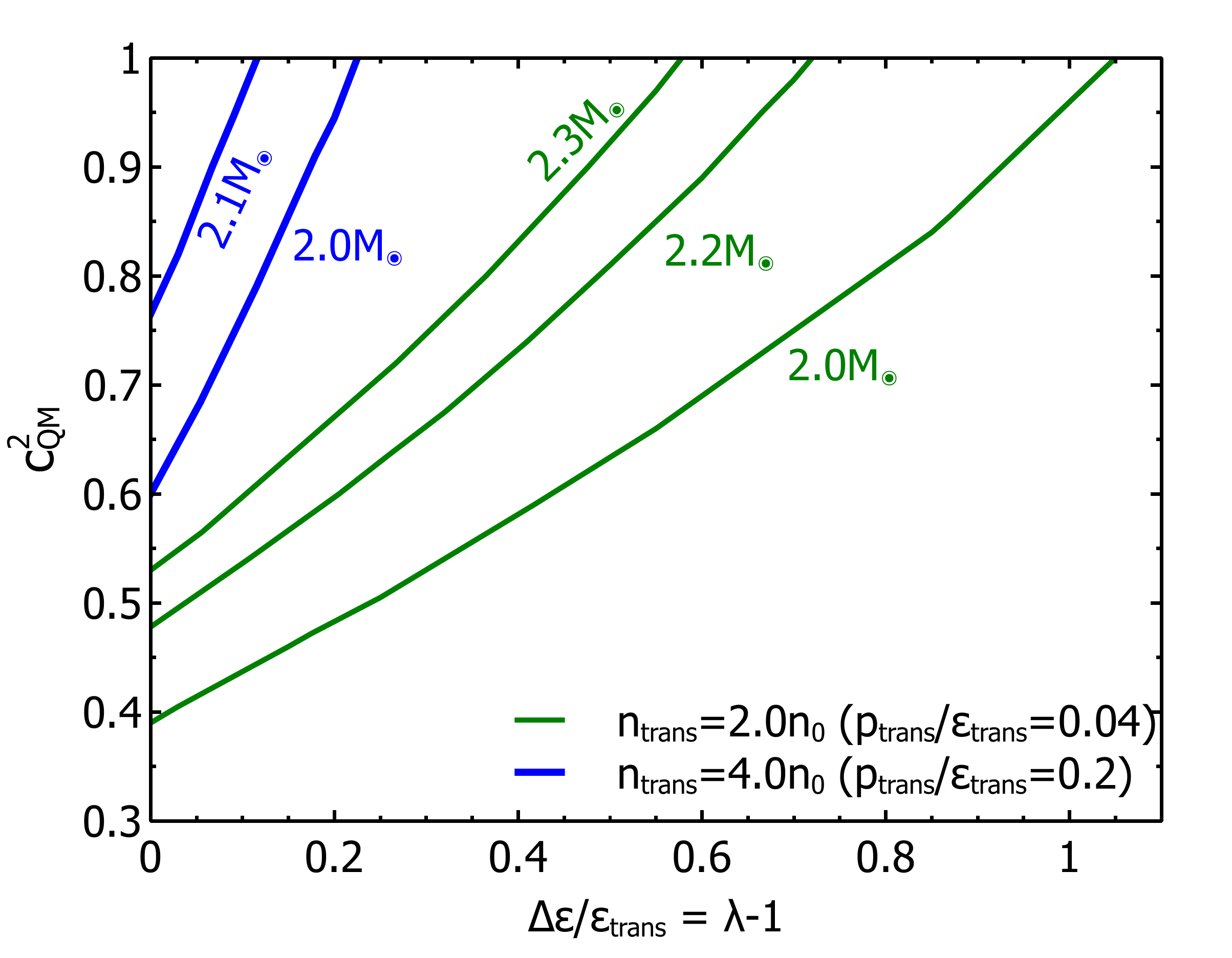}}
{\includegraphics[width=200pt,angle=0]{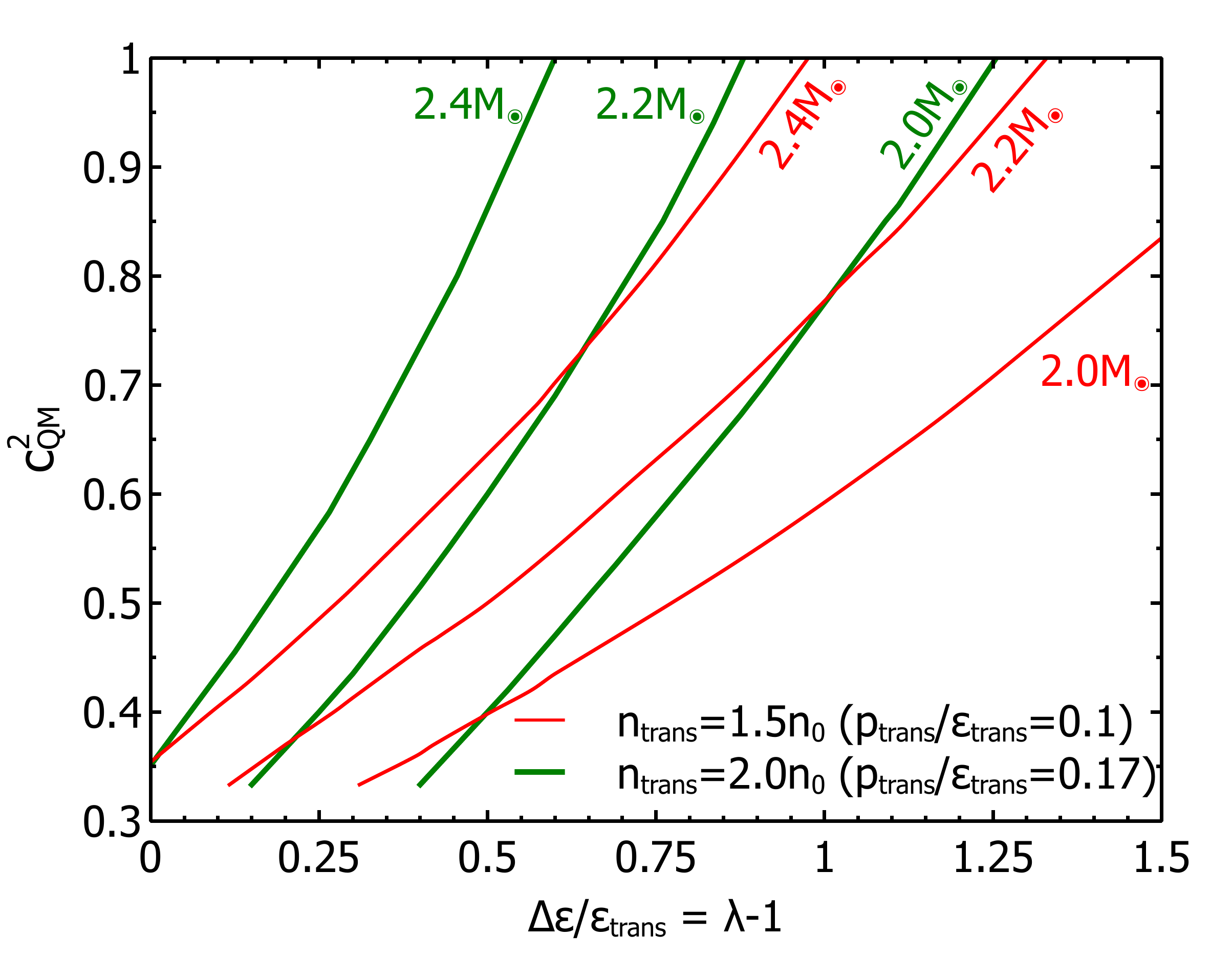}}
\caption{(Color online) Mass of the heaviest hybrid star vs quark matter EOS parameters $p_{trans}/\epsilon_{trans}$, $c_{QM}^2$, and $\Delta\epsilon/\epsilon_{trans}$ 
for HLPS (left panel) and NL3 (right panel) nuclear matter. The thin (red), medium (green) and thick (blue) lines are for nuclear to quark transition at 
$n_{trans}=1.5n_0,~2n_0$ and $4n_0$, respectively. Figure adapted from Ref. \cite{Alford13}.}
\label{maxmasses}
\end{figure}
%

Figure \ref{maxmasses} shows two illustrative examples for $\epsilon_{{\rm NM}}(p)$: a 
relativistic mean field model labeled NL3 \cite{Shen:2011kr} 
and a non-relativistic potential model labeled HLPS,
corresponding to ``EoS1'' in Ref.~\cite{Hebeler:2010jx}.  The EOS of HLPS is softer than that of NL3 at low density, so the contrasts between these EOS's are apparent in the results. 
The main lesson learned is that it is possible to get  hybrid stars in excess of 2 M$_\odot$ for reasonable parameters of the quark matter EOS:  
not-too-high transition density ($n\sim 2n_0$), low enough energy density discontinuity $\Delta\epsilon < 0.5~\epsilon_{trans}$, and high enough speed of sound $c_{QM}^2 \geq 0.4$. 

\begin{figure}[thb]
\begin{center}
\includegraphics[width=250pt]{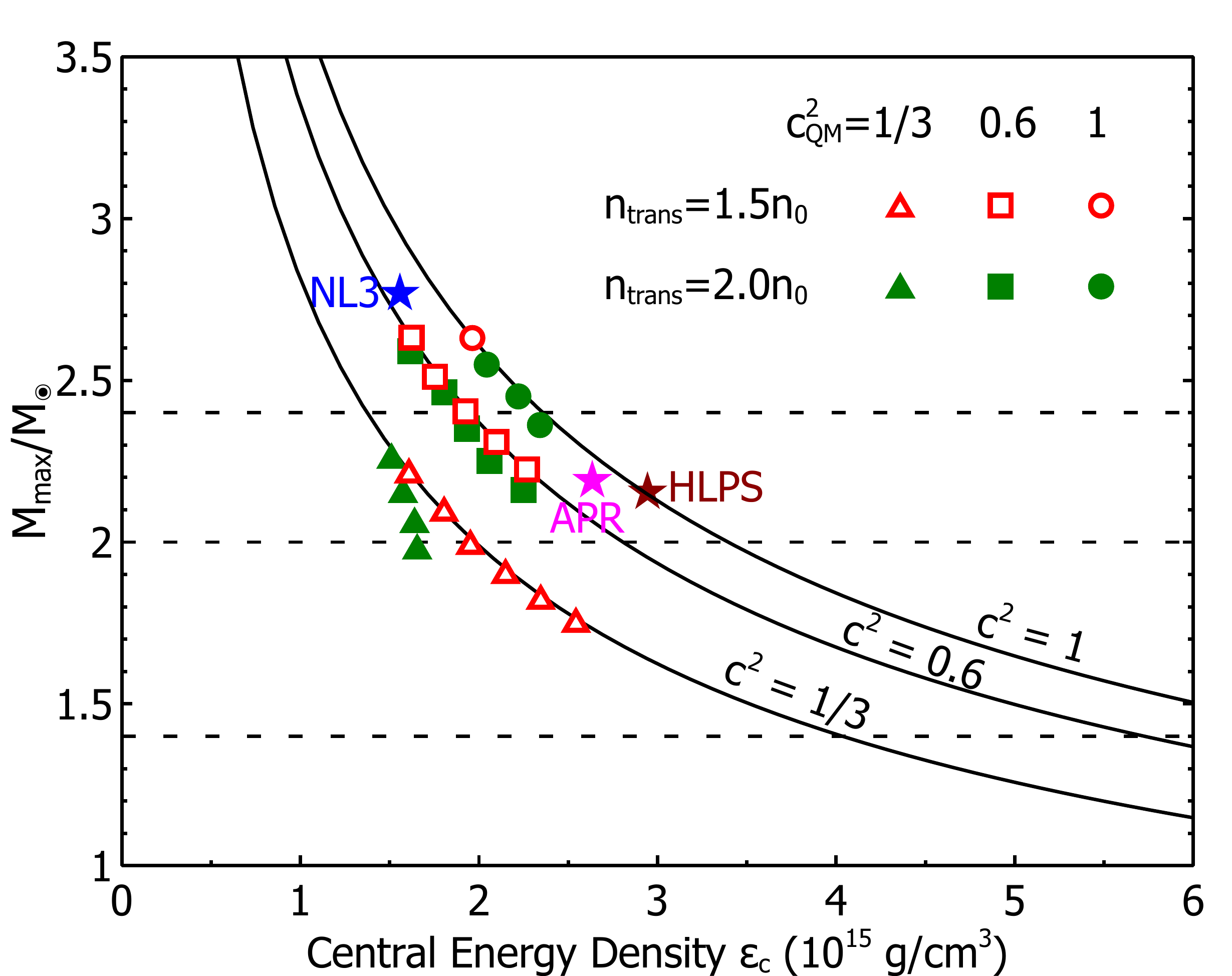}
\caption{ (Color online) Mass of the heaviest hybrid star vs. its central 
energy density, for various quark matter EOS's in equation (\ref{eqn:EoSqm1}).
The curves are 
predictions of Ref.~\cite{LP:11} for 
stars whose core-region squared speed of sound is 1. 0.6, and 1/3.  
Pure nuclear matter stars for the NL3, HLPS  and APR \cite{APR} equations of state
are also plotted. Figure adapted from Ref. \cite{Alford13}.}
\label{fig:Mm-ec}
\end{center}
\end{figure}

The finding in Refs.~\cite{LP:11,LP:05} that $M_{max} \propto \epsilon_{cent}^{-1/2}$ is borne out for hybrid stars as well (see Figure \ref{fig:Mm-ec}). For the low transition density, $n_{trans}=1.5n_0$ and $2n_0$ cases shown, hybrid stars have large quark matter cores which give  substantial contributions to the maximum masses.  
Perturbative treatments are characterized by $c_{QM}^2\simeq1/3$, and a value of $c_{QM}^2$ above 1/3 indicates that  quark matter, if present in neutron stars, is strongly coupled. Non-perturbative treatments of quark matter are  sorely needed.  

Since the discovery of $2~{\rm M}_\odot$ stars, many efforts have been made in the literature (a search in the arXiv is recommended) to verify the extent to which hyperons could be present in neutron stars.   With modifications of earlier treatments, some succeed, but others don't, in producing the required mass. Clearly work will continue along these lines, which is necessary as it will also serve to benefit the treatment of interactions involving strangeness-bearing particles in relativistic heavy-ion collisions at RHIC and LHC.  

In summary about the maximum mass, 
the larger the observed neutron star mass,  larger is the aggravation for nuclear theory to deliver a believable EOS that can support it. Its importance extends to the minimum mass of a black hole and the total number of stellar mass black holes in our Universe (of concern to Cosmology), and the progenitor mass (of concern to stellar evolution) besides the EOS of dense matter.   

The minimum mass of a neutron star is no less important than its maximum mass. Well measured minimum masses hover around  $1~{\rm M}_\odot$ (see the figure of measured neutron star masses in Ref. \cite{Lat12}). In the current paradigm of neutron star formation in the aftermath of a core collapse supernova, only progenitor stars with core masses around the Chandrasekhar mass limit of $\sim 1.4~{\rm M}_\odot$ are  suitable candidates.
How then could a low mass neutron star be produced from stellar evolution theory? See the review by Lattimer \cite{Lat12} in which the current conundrum in this regard is detailed. 

\subsection{Masses and radii of individual neutron stars}
\label{sec:radii}
Radio pulsar measurements, precise for neutron star mass measurements when general relativistic effects can be measured, cannot yet yield radius information of the same neutron star for which a mass has been well measured. Possibilities exist in a double neutron star binary through moment of inertia measurements that exploit effects of spin-orbit coupling \cite{LS:2005}, but await developments in accurate pulse timing techniques.  Precise measurements of masses and radii of several individual neutron stars would pin down  the EOS of neutron star matter without recourse to models  \cite{Lindblom:92,Prakash:09}. 

The importance of simultaneous mass and radius measurements has been realized by several workers.  Lattimer has reviewed the current status in Ref. \cite{Lat12} with extensive references.   The news here is that (1) methods are being devised to make simultaneous measurements of $M$ and $R$ possible, and (2) observers and theorists are working, jointly or separately, to reach agreement in the methods of analyses and conclusions.  Directions in which significant efforts have been made include observations and analyses of X-ray emission from
\begin{enumerate}
\item isolated neutron stars,
\item intermittently quiescent neutron stars undergoing accretion from a companion star, and
\item neutron stars that display type I X-ray bursts from their surfaces.
\end{enumerate}
\subsection*{Isolated neutron stars}
%
\begin{table}[hbt]
\begin{center}
\begin{tabular}{|c|c|c|c|} 
\hline
Isolated Neutron Star & $T_\infty$ & $P$ & $D$    \\ 
             & (eV) & (s) & (pc) \\
\hline
RX J0420.0-5022 & 44 & 3.45 & $\cdots$ \\
RX J0720.4-3125 & 85-95 & 8.39 & $330^{+170}_{-80}$ \\
RX J0806.4-4123 & 96 & 11.37 & $\cdots$ \\
RX J1308.8+2127 & 86 & 10.31 &$\cdots$  \\
RX J1605.3+3249 & 96 & 6.88? & $\cdots$ \\
RX J1856.5-3754 & 62 & 7.06 & $120\pm8$ \\
RX J2143.0+0654 & 102 & 9.44 & $\cdots$\\
\hline
\end{tabular}
\caption[RX J's]{The ``magnificent seven'' isolated neutron stars and their key features. 
Spectral analysis yields the temperature $T_\infty$. 
X-ray pulsations (these stars are radio-quiet) are used to infer the spin period $P$. 
Parallax and proper motion motion measurements have allowed the distance $D$ to be well determined only for RX J1856.5-3754 \cite{Walter2002}.  Entries in this table are extracted from Refs. \cite{Lat12,Kaplan08}.}
\label{magnificent7}
\end{center}
\end{table}
%
Seven isolated neutron stars - referred to as the ``magnificent seven'' -  have received much attention through  {\em Chandra, XMM} and {\em HST} observatories since their discovery in the all-sky search by the {\em Rosat} observatory. Some properties relevant for the determination of $M$'s and $R$'s of these objects are listed in Table \ref{magnificent7}. 

The modus operandi to simultaneously infer $M$ and $R$ proceeds as follows.  The observed flux (for the most part in X-rays except when the object is nearby when optical data is also available) is fit using 
\begin{equation}
F = 4\pi \sigma~ T_\infty^4 \left(\frac {R_\infty}{D} \right)^2  \qquad {\rm and} \qquad 
R_\infty = R~\left( 1 - \frac {2GM}{c^2R} \right)^{-1/2} \,,
\end{equation}
where $\sigma$ is Boltzmann's constant. Above, symbols with subscripts $\infty$ refer to an observer at a far distance from the source. The temperature fit to the data by the distant observer is    $T_\infty = T \left[ 1-2GM/(c^2R)\right]^{1/2}$, where $T$ is the temperature at the source.  The ``radiation radius'',  $R_\infty$, depends on $M$ and $R$ as indicated above. When the distance to the star, $D$, is well known (a rare occurrence  as Table \ref{magnificent7} shows), the radius and the mass of the star can be determined using 
\begin{equation}
R = R_\infty ~(1+z)^{-1} \qquad {\rm and} \qquad 
\frac {M}{\rm {M}_\odot} = \frac {c^2R}{2G{\rm M}_\odot} ~ \left[1-(1+z)^{-2}\right] \,, 
\end{equation}
by treating $T_\infty$, $R_\infty$ and $z =  \left[ 1-2GM/(c^2R)\right]^{-1/2} - 1$ 
(the surface redshift factor) as parameters in the spectral analysis.  Straightforward as it seems, real life bites on several fronts!
A thermally emitting neutron star is not a perfect blackbody as its atmospheric composition, and the strength and structure of its magnetic field (the periods in Table \ref{magnificent7} give some indications) are unknown. The interstellar hydrogen absorption (generally taken from independent sources) is an additional parameter. The situation with the often studied case, RX J1856.5-3754, is summarized in Table \ref{stonybrookstar}.
\begin{table}[hbt]
\begin{center}
\begin{tabular}{|c|c|c|c|l|l| }
\hline
 $R_\infty$ (km) & $z$ & $R$ (km) & $M~({\rm M}_\odot) $ & Atmospheric model & Ref.   \\ 
\hline
 $16.1\pm1.8$ & $0.37\pm0.03$ & $11.7\pm1.3$ & $1.86\pm0.23$ & Non-magnetic heavy elements & \cite{Pons02} \\  
 $\simeq 15.8$ & $\simeq 0.3$ & $\simeq 12.2$ & $\simeq 1.68$ & Non-magnetic heavy elements & \cite{Walter04} \\ 
 $>13$ & $\cdots$  & $\cdots$ &$\cdots$ &Condensed magnetized surface & \cite{Burwitz03} \\
 $14.6\pm1$ & $\simeq0.22$ & $11.9\pm0.8$ & $1.33\pm0.09$ & Condensed magnetized surface;  & \cite{Ho07} \\
 & & & & trace H & \\
\hline
\end{tabular}
\caption[RX J1856-3754]{Inferred radius and mass of the isolated neutron star RX J1856.5-3754 from different atmospheric models using data from the {\em Rosat, HST, Chandra} and {\em XMM} observatories.}
\label{stonybrookstar}   
\end{center}
\end{table}

Although the inferred radii are similar, the masses are not. Not happy news! The magnificence of the seven is yet to be revealed!

\subsection*{Quiescent neutron stars}
\begin{figure}[thb]
\begin{center}
\includegraphics[width=\hsize]{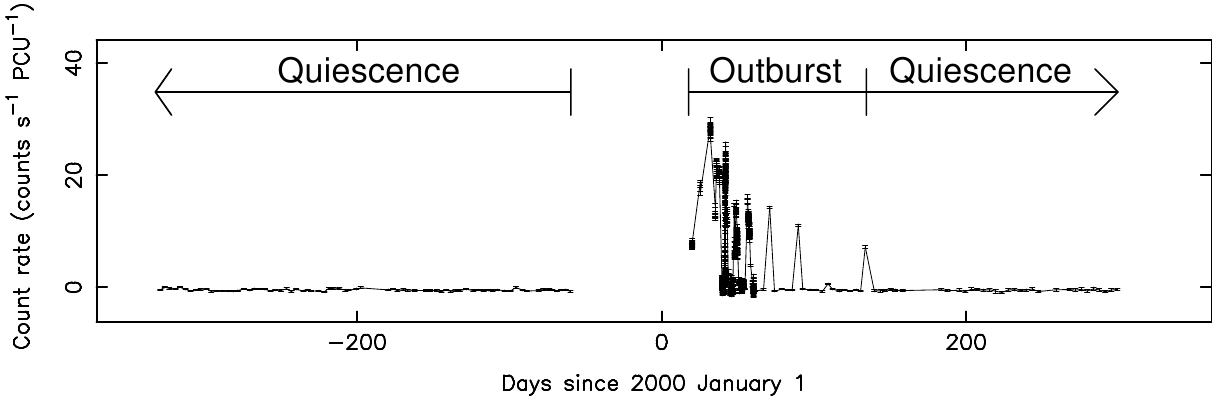}
\caption{(Color online) Quiescent periods of a neutron star following X-ray burst activity. Figure courtesy Ed Cackett.} 
\label{fig:burst}
\end{center}
\end{figure}
\begin{figure}[h]
\begin{center}
\includegraphics[width=220pt]{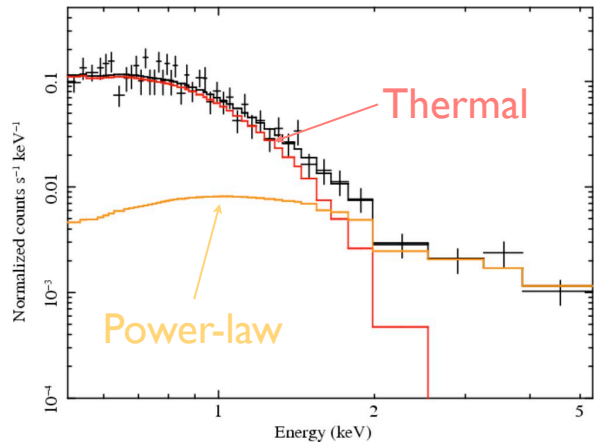}
\caption{(Color online) Thermal emission during quiescent periods following X-ray burst activity. Figure courtesy Ed Cackett.} 
\label{fig:xrays}
\end{center}
\end{figure}
Many neutron stars go through long periods of quiescence between episodes of intermittent accretion from a companion star (see Fig. \ref{fig:burst} for an example).  Pycno-nuclear reactions, triggered by the compression of matter in the crust caused by accretion,  release energy which heats the crust \cite{haensel90}. In the quiescent periods, the heated crust cools via thermal emission from the surface detectable as X-rays \cite{Brown98}  as shown in Fig. \ref{fig:xrays}.  The observed spectra are generally fit with well-understood nonmagnetic H atmospheres as lack of pulsations or cyclotron frequencies indicate insignificant magnetic fields.  References \cite{Heinke06,Webb07,Guillot11} describe analyses to infer the apparent angular area and the surface gravity. Figure 10 of Ref. \cite{Lat12} shows probability distributions of $M$ and $R$ for neutron stars in globular clusters M13, X7, $\omega$ Cen and U24 constructed by Andrew Steiner.  For the cases shown there, the $M$ are $R$ values inferred are not very restrictive.

\subsection*{Type I X-ray bursts undergoing photospheric radius expansion}
Sustained accretion can cause the envelope of a neutron star to become thermally unstable to He or H ignition which leads to a thermonuclear explosion observed as an X-ray burst with a rapid rise time ($\sim$ 1 s) followed by a cooling stage lasting  $\sim$ 10-100 s \cite{Stroh04}.  
When the bursts are sufficiently luminous, the surface layers of the neutron star and the photosphere are driven outward to larger radii by the radiation pressure which can match or even surpass  the Eddington value for which the radiation pressure balances gravity. 
In Refs. \cite{Ozel12,Steiner10,Steiner13,Suleimanov11} the bursters EXO 1745-248, 4U 1608-522, 4U 1820-30 and KS 1731
have been modeled to simultaneously infer the masses and radii of neutron stars.  
Physical parameters of these models include the opacity of the lifted material, the effective blackbody temperature when the lifted material falls down to the 
surface after expansion (touchdown), the color correction factor that accounts for effects of the atmosphere in distorting the inferred temperature, possible models of atmospheres, and whether or not the photosphere radius is equal to or larger than the radius of the neutron star.  
The situation is far from settled as
the inferred values of radii have ranged from 8-10 km \cite{Ozel12}, 11-13 km \cite{Steiner10,Steiner13} and in excess of 14 km \cite{Suleimanov11}. Efforts are in progress to achieve consistency with data on nuclei (for summaries, see, e.g., \cite{Lat12,Steiner05}) and those from heavy-ion collisions in which the collective flow of matter, momentum and energy have been analyzed \cite{Danielewicz02}. 

\subsection{Cooling of the neutron star in Cassiopeia A}
\label{sec:CasA}
The neutron star in Cassiopeia~A (``Cas~A" hereafter) discovered
in 1999 in the {\em Chandra} first light observation
targeting the supernova remnant, 
is one of the youngest known in the Milky Way.  An association
with the historical supernova SN~1680 places 
Cas~A at an age of 333 yrs. 
The distance to the remnant is $\simeq 3.4$~kpc.  
The observed thermal soft X-ray spectrum of
Cas~A implies 
a surface temperature $\cong 2\times 10^6$~K and an
emitting radius $\sim 8 - 17$~km.  These results raise Cas A to
the rank of one of the very few isolated neutron stars with a
well determined age and a reliable surface temperature, thus allowing
for a detailed modeling of its thermal evolution and the determination
of its interior properties.  
\begin{figure}[htb]
\begin{center}
\includegraphics[width = \hsize,angle=0]{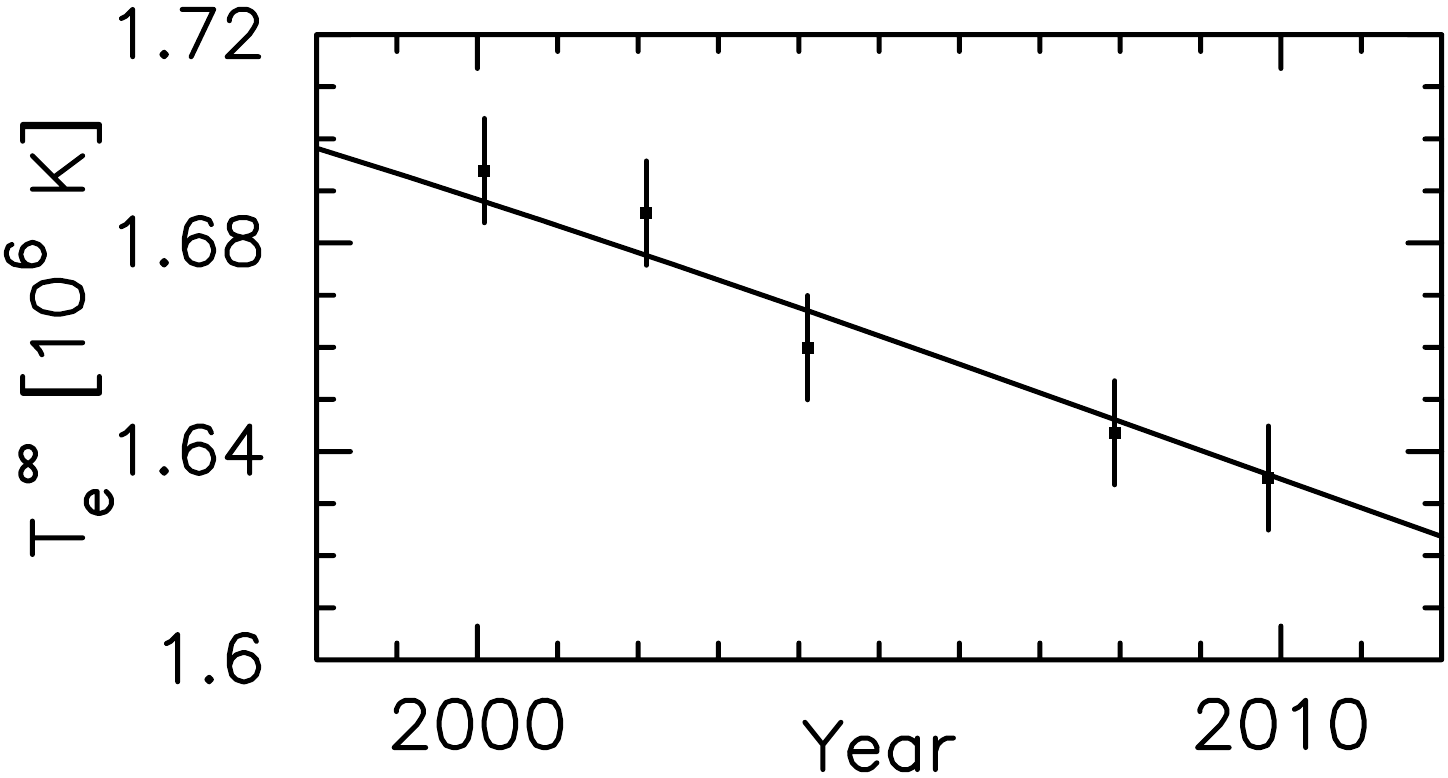}
\end{center}
\caption{Red-shifted surface temperature of Cas A reported by Heinke and Ho \cite{HeHo10}.  
Figure courtesy D. Page.}
\label{CasA}
\end{figure}

\begin{figure}[htb]
\begin{center}
\includegraphics[width = 0.85\textwidth,angle=0]{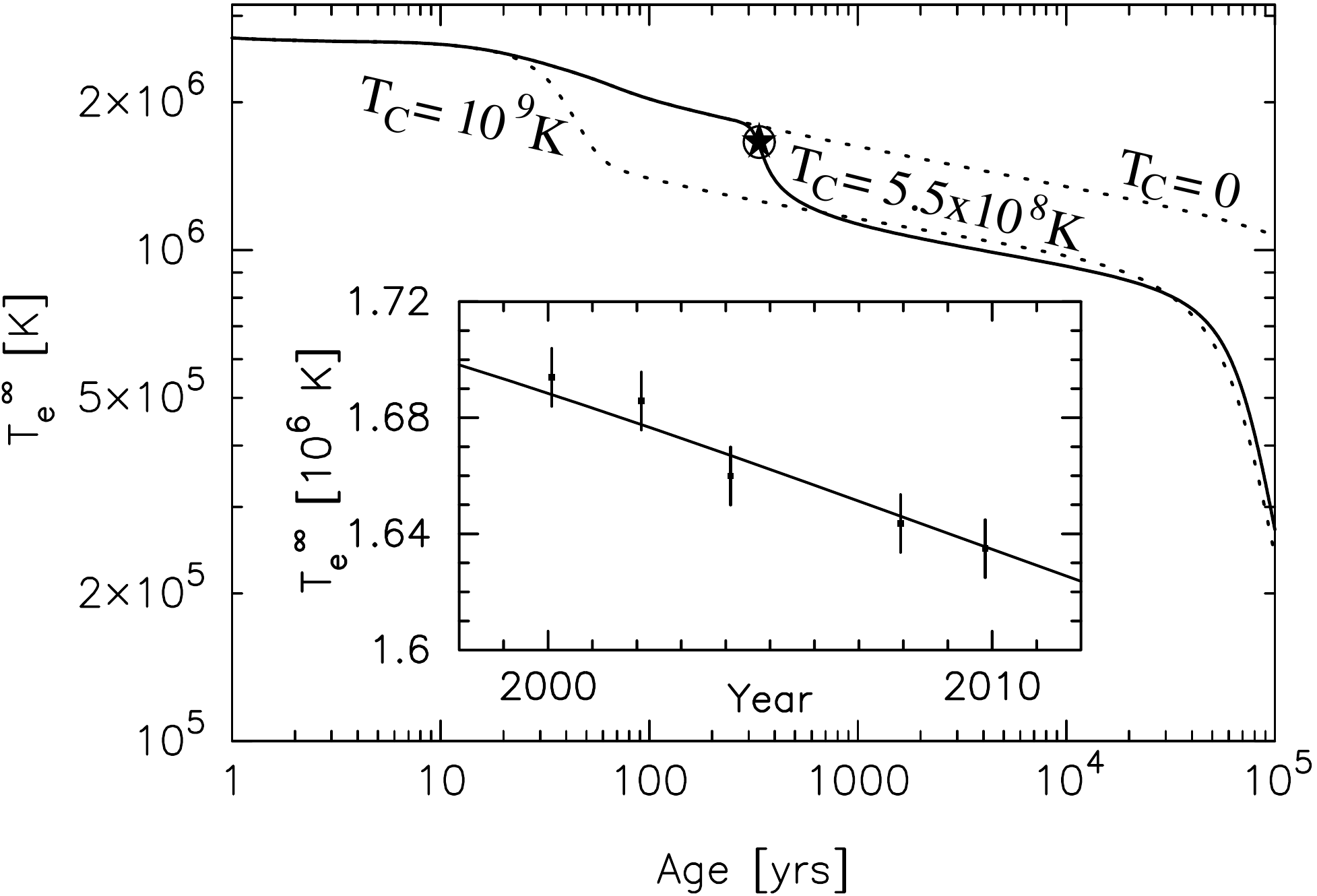}
\end{center}
\caption{Critical temperature $T_c$ for neutron superfluidity deduced by detailed modeling (see Ref. \cite{Page10}  for details) of Cas A's cooling from its birth. Figure courtesy D. Page.}
\label{goodfit}
\end{figure}

Analyzing 10 years (from 2000 to 2009) of archival data, Heinke and Ho~\cite{HeHo10}
recently reported that Cas~A's surface
temperature has rapidly decreased from $2.12 \times 10^6$ to $2.04
\times 10^6$~K (see Fig. \ref{CasA} for red-shifted surface temperature).  This reported rate of cooling is significantly
larger than that from the standard neutrino emitting processes 
expected to occur in the core of a neutron star. Although many possibilities for rapid cooling exist, the rapidity with which cooling has occurred in Cas A at such an early stage of its life sheds light on the most likely process. A successful accounting of the observed data involves superfluidity of neutrons and superconductivity of protons in the core of Cas A.  As the star is cooling down due to neutrino emission processes, a stage is encountered when the temperature falls below the critical temperature for neutron and  proton pairing phase transitions  leading to the phenomena of superfluidity and superconductivity. At temperatures just below the critical temperature of a pairing transition, the continuous breaking and formation of pairs results in enhanced neutrino emission (see, e.g., \cite{Page04,Page09} and references therein). 
Numerical calculations and analytical analysis 
performed by Page, Prakash, Lattimer and Steiner (all Stony Brookers) \cite{Page10}
indicated
a critical temperature $\simeq 0.5\times 10^9$ K for the triplet
neutron superfluidity. 
The observed rapidity of the cooling implied that protons were already in a superconducting state with a larger critical temperature. See Fig. \ref{goodfit} for a good fit to Cas  A's current temperature and cooling rate \cite{Page10}. 

The reported rapid cooling of Cas A was the first direct evidence that superfluidity and superconductivity occur at supra-nuclear densities in the cores of neutron stars. 
Two days after the work of Page et al. was submitted for publication, Shternin et al., \cite{Shternin11} 
arrived at similar conclusions in an independent work.  Since these works, an alternative scenario involving in-medium modifications to neutrino emission processes that could explain the observed features (without superfluidity playing a major role) has been reported in Ref. \cite{Blaschke11}. 

The possibility of degradation of detectors aboard {\em Chandra} and associated modifications in calibrations raised by Rutledge (private communication) 
were then addressed in the work of Elshamouty et al., \cite{Elshamouty13} in which data from all  relevant  {\em Chandra} detectors were analyzed.  
Combining the data from these detectors, the best estimate of decline in the surface temperature over 10 years was quoted to be $2.9\pm0.9~ ({\rm stat.})^{+1.6}_{-0.3}~({\rm sys.})\%$ in contrast to the original $\sim 4\%$ decline in \cite{HeHo10}. 
\begin{figure}[htb]
\begin{center}
{\includegraphics[width=\hsize,angle=0]{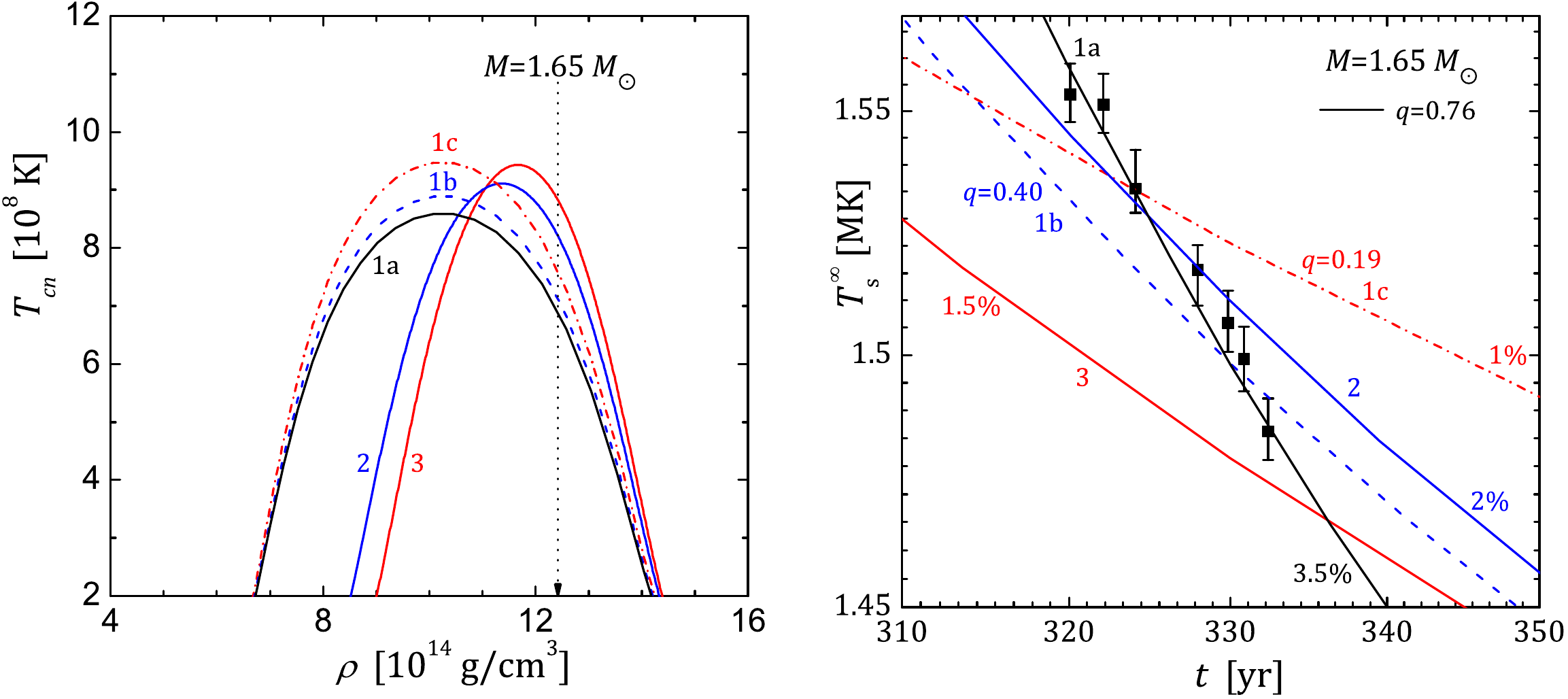}}
\end{center}
\caption{(Color online) Left panel:  Models of the critical temperature for triplet-neutron pairing vs baryon density explored in  Ref. \cite{Elshamouty13} for a $1.65~{\rm M}_\odot$ neutron star. Right panel:  The surface temperature in $10^6$ K  at a far distance from the source vs age for the possibilities of superfluidity in the left panel ($q$ is the reduction factor for the PBF $\nu$-emission by many-body effects). The data shown are for the largest decline in the surface temperature. The various curves are for theoretical expectations for different levels of decline observed in the detectors aboard {\em Chandra}.  
Figure adapted from Ref. \cite{Elshamouty13}. }
\label{CoolCurves}
\end{figure}

The theoretical implications of this new data were examined in  Ref. \cite{Elshamouty13} and are shown in Fig. \ref{CoolCurves}. Earlier conclusions about the role of neutron superfluidity that sets the surface temperature of   $\cong 2\times 10^6$~K were confirmed. The rate of decline in the temperature is chiefly set by proton superconductivity, and concerns about the uncertain and extreme models used earlier have been much abated, to the relief of Page et al., and possibly also to that of Shternin et al..  
 Analysis of additional data by an independent group \cite{Posselt13} indicates next to nil cooling over 10 years, and has resulted
in a mild controversy.  Reference  \cite{Elshamouty13} claims cooling is
likely present, albeit at a lower rate than reported initially.  In contrast, \cite{Posselt13}  maintains that cooling is likely
not present and attributes the cooling reported in \cite{Elshamouty13} as due to instrumental effects and remnant environmental changes. Both these reports appear to be consistent with cooling, but to different degrees of significance which remains to be
sorted out.

In summary about Cas A, we are fortunate in being able to record its thermal emission continually with {\em Chandra}.   As long as it is able, {\em Chandra} should be  made available for observing programs until a better observatory is in place. Much can be learned about ``Stellar Superfluids'' as emphasized in the book chapter by Page et al. \cite{Page13} in the monograph ``Modern Superfluids''.

\subsection{Crustal cooling of neutron stars}
\label{sec:Crustalcooling}
\begin{figure}[htb]
\begin{center}
{\includegraphics[width=0.65\textwidth,angle=0]{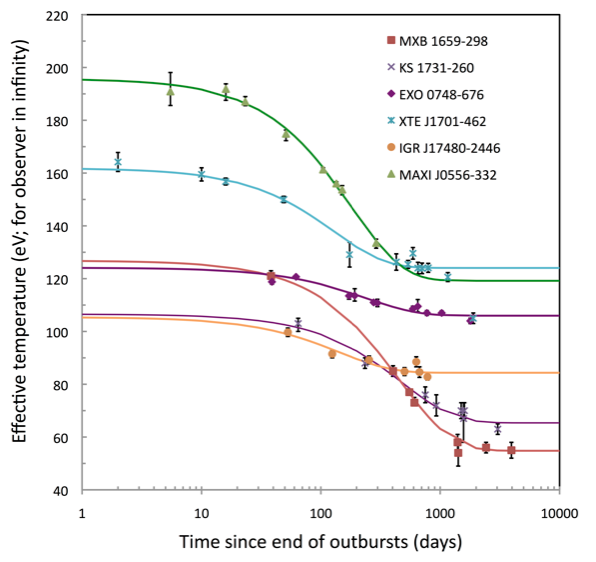}}
\end{center}
\caption{Cooling curves following X-ray bursts. Figure courtesy Ed  Cackett.}
\label{crustscool}
\end{figure}

Mal Ruderman was the first to suggest that neutron stars would have solid crusts \cite{Ruderman68} (Ed Cackett dug this information out at my request).  It's remarkable that they indeed do, as the cooling curves of several neutron stars after X-ray bursts following accretion indicate (see Fig. \ref{crustscool}).  Pertinent observations can be found in Refs. \cite{Cackett,Fridriksson}.
Recent theoretical interpretation of these cooling curves can be found in Page and Reddy \cite{Page13b}, who analyze the time scales for cooling in terms of the amount of mass accreted, specific heat, thermal conductivity, nuclear heating and neutrino cooling rates at the relevant layers. 
   
\section{Puzzles}
\label{sec:puzzles}
Several  puzzles require solutions in the physics and astrophysics of neutron stars. I list below a few puzzles that bother many including   me.
\subsection{A neutron star or a black hole in SN 1987A?}
There is no news of a neutron star in SN 1987A. How long does it take for a neutron star to be revealed after a core-collapse supernova explosion?  Did SN 1987A end up as a black hole?  If so, was it upon fall-back accretion (Brown - Bethe scenario \cite{BB94}), or,  after the deleptonization stage (Prakash - Lattimer scenario \cite{ELP96})?

\subsection{Evidence for enhanced cooling?}
\begin{figure}[htb]
\begin{center}
\includegraphics[width=0.55\textwidth,angle=270]{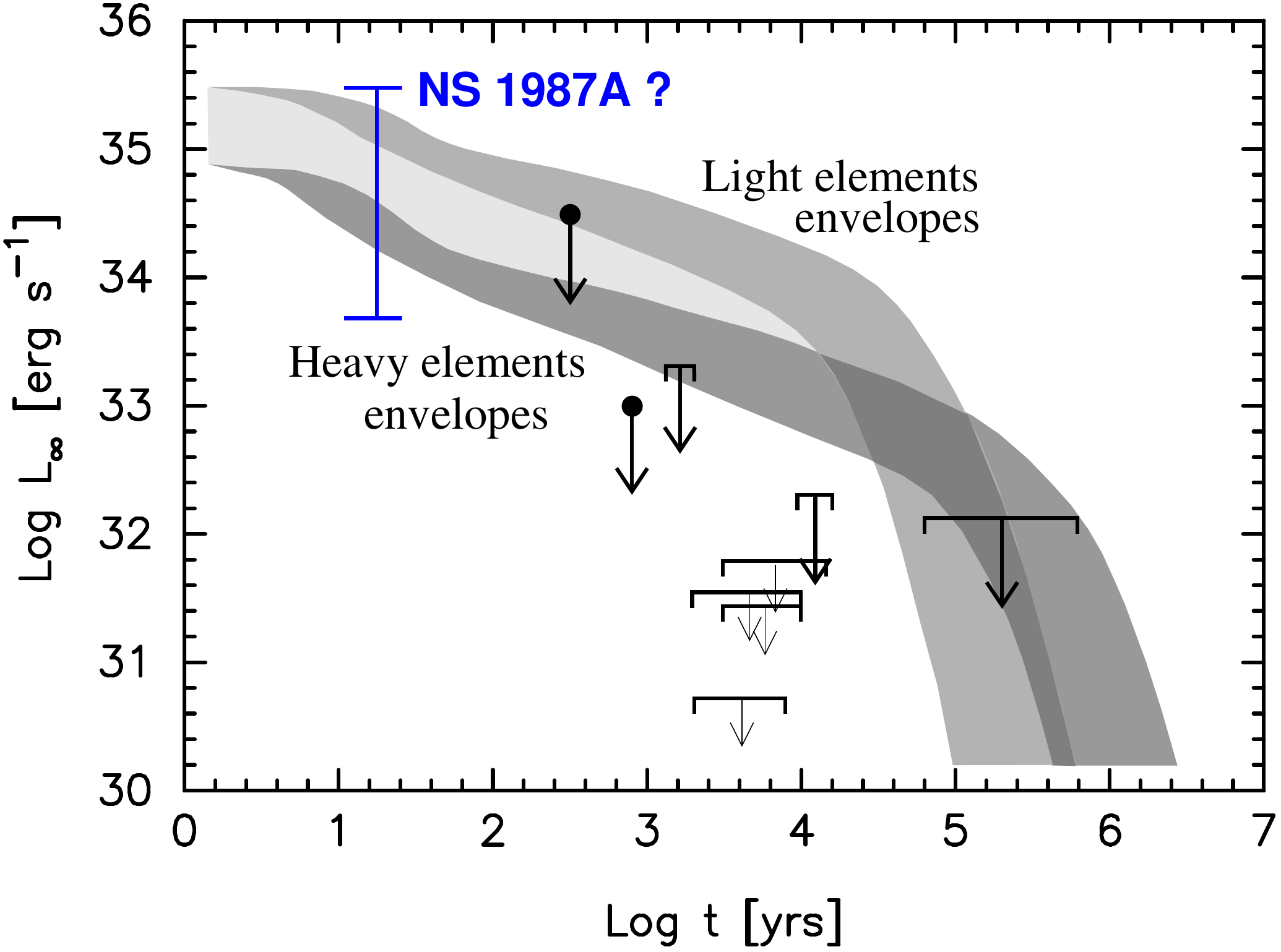}
\end{center}
\caption{(Color online) Really cold isolated objects. Upper limits of the data shown are from Ref. \cite{Kaplan04}. The shaded bands are 
model predictions of the ``Minimal Cooling'' paradigm \cite{Page04}.}
\label{coldies}
\end{figure}

In order to account for the measured cooling curves of neutron stars, 
the ``Minimal Cooling'' paradigm excludes {\em a priori}
all possible fast $\nu$-emission mechanisms, thus
restricting $\nu$-emission to the ``standard" MU process and the
similar nucleon bremsstrahlung processes~\cite{Page04}.  
However, effects of nucleon pairing, i.e., neutron
superfluidity and/or proton superconductivity, are included. 
While successful in explaining the data for the most part,  
several cases \cite{Kaplan04} fall below the ``Minimal Cooling'' paradigm (see Fig. \ref{coldies}) and
point to enhanced cooling, if these objects correspond to neutron stars. Do these objects contain neutron stars or black holes?

\subsection{Why don't neutron stars spin up to the theoretically allowed limit?}
\begin{figure}[htb]
\begin{center}
\includegraphics[width=0.99\textwidth,angle=0]{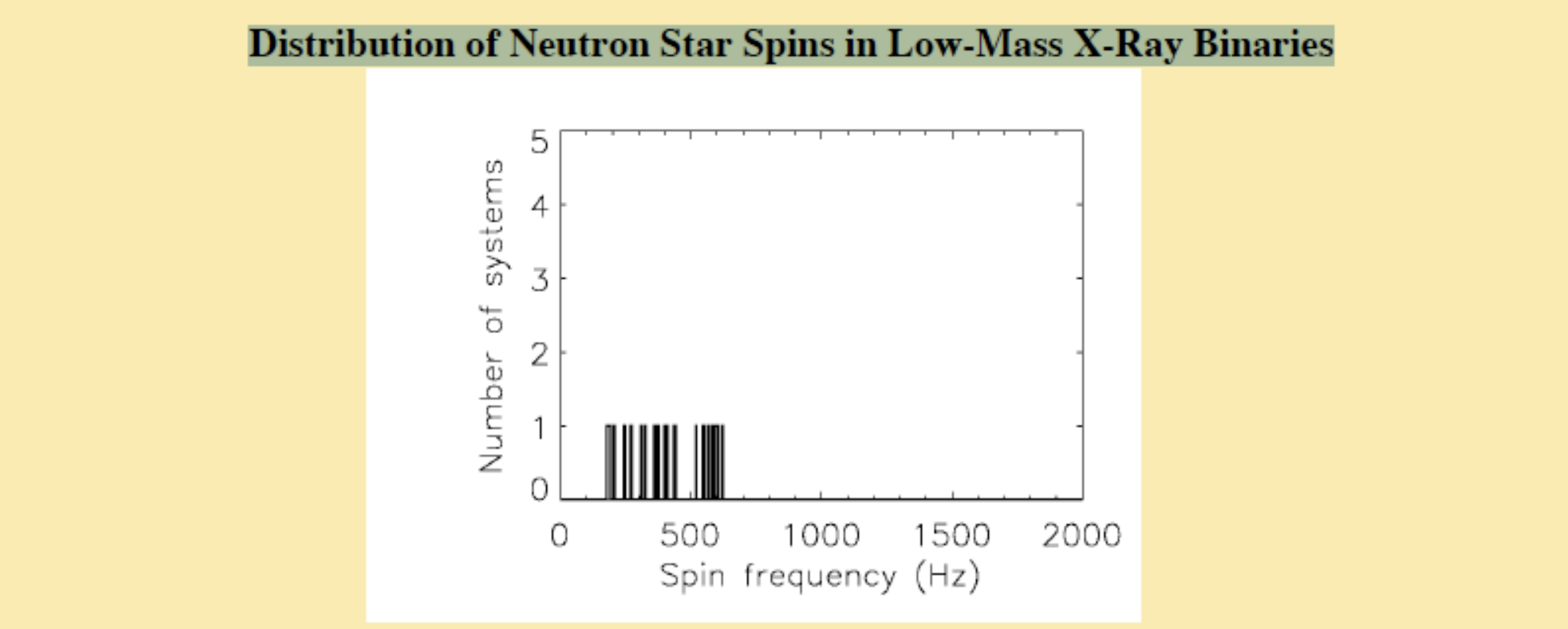}
\end{center}
\caption{(Color online) Distribution of neutron star spin frequencies \cite{Deepto}. Figure courtesy Deepto Chakraborty. }
\label{spins}
\end{figure}

How fast can neutron stars spin? General relativistic calculations of rapid rotation give an upper limit to the neutron star spin frequency \cite{Rots,LP04}
\begin{eqnarray}
\nu_{max}(M_{max}) &\simeq& 1224~\left(\frac {M_{max}}{\rm M_\odot}\right)^{1/2} 
\left(\frac{R_{max}}{10~{\rm km}}\right)^{-3/2} ~{\rm Hz} \,, \nonumber \\
\nu_{max}(M)&\simeq&1045 ~\left(\frac {M}{\rm M_\odot}\right)^{1/2} 
\left(\frac{R}{10~{\rm km}}\right)^{-3/2} ~{\rm Hz} \,, 
\end{eqnarray}
where $M_{max}$ and $R_{max}$ are the non-rotating maximum mass spherical configurations, and the second relation refers to a mass not too close to the maximum mass. Effects of general relativity being  the largest for  the most compact configurations (largest $M/R$ values), sub-millisecond rotation is of much interest.  The observed distribution of neutron star spins in low-mass X-ray binaries is shown in Fig. \ref{spins}. The puzzle is, there is a sharp cutoff for spins above 730 Hz well below the theoretically allowed upper limit. Can gravitational radiation from rapidly rotating pulsars undergoing accretion limit their spin up? This possibility is being explored by many in the field.   

\subsection{Where and how are the heavy elements made?}
The sites and the manner in which elements heavier than iron are made in our Universe have been  longstanding puzzles. Candidate sites are  the far suburbs of  where core-collapse supernova explosions and neutron star mergers occur.  Conditions suitable for rapid neutron capture processes to occur in the aftermath of a supernova explosion are yet to be realized in simulations. The recent observations of a kilonova event \cite {Tanvir13,Berger13} in the short gamma-ray burst of GRB 130603B 
have  rejuvenated interest in binary neutron star mergers  in the context of heavy element synthesis. Lattimer's thesis work in the mid 1970's \cite{Lat76} laid the ground for thinking about binary star mergers as a site for synthesizing heavy elements.  I hope he gets rewarded when this puzzle is solved!     

\section{My tribute to Gerry}
\label{sec:tribute}
Tusind tak! I miss you! A lot!! 

Love and regards from Manju, Ellen and, particularly, Smita. 

For those of you at this meeting who saw the video of the party at Edward Shuryak's house, the little girl sitting on Gerry's lap is my daughter Smita. As things turned out, Smita nursed him while he was seriously ill. It was amazing that Betty and Smita were the only two nurses Gerry felt comfortable with when he was at the hospital. 

\section*{Acknowledgments} 
I thank all the PALS  I got to know through Gerry. 
This research is supported in part by the US DOE under Grant No. DE-FG02-93ER-40576.  
%
%

\end{document}